# PRECESSIONS IN RELATIVITY

COSTANTINO SIGISMONDI

*University of Rome "La Sapienza" Physics dept. and ICRA,
Piazzale A. Moro 5 00185 Rome, Italy. e-mail: sigismondi@icra.it*

From Mercury's perihelion precession (open question from 1845 to 1915) to Gravity Probe-B satellite (first proposal in 1959, launch in 2004), General Relativity had to deal with precession phenomena. Perihelion advancement precession (Einstein), geodetic (de Sitter), gravitomagnetic (Lense-Thirring) and spin-spin (Pugh-Schiff) precession are compared with all Newtonian terms in cases of weak and strong gravitational fields. Qualitative models and order of magnitude estimates from classical papers are presented.

## 1. Mercury's perihelion precession

In General Relativity the properties of planetary orbits are dependant on spacetime structure. In Schwarzschild metric the two-body problem yields rosette orbits, with an advancing perihelion. For Mercury the rate of advancement is $\delta\theta \approx 43$ arcsec/cy, and Einstein in 1915 [1] showed that $\delta\theta = 6\pi GM_{sun} \cdot a/c^2 b^2$ with a,b semiaxes of ellipse $b=a\sqrt{1-e^2}$, e= eccentricity. For other planets or satellites $\delta\theta \sim M/R$, with M mass of the central body and R radius of the orbit supposed circular. For the Moon this yields an effect 1600 times smaller than for Mercury. In special relativity there is also an effect of perihelion advancement, its magnitude can be derived considering the time dilation formula $t=t_0/\sqrt{1-v^2/c^2}= t_0 \cdot \gamma$ at fixed reference frame of the central mass, source of Schwarzschild field. Each $\delta\ell$ of Newtonian elliptical orbit occurs in a locally inertial frame and special relativity applies to. The whole orbit lasts an extra time $\Delta t/T = \gamma-1$. From the virial theorem in a circular orbit $v^2=GM/R$, expanding in Taylor series $\gamma \approx 1+\frac{1}{2}GM/Rc^2$, so $\Delta t/T = \frac{1}{2}GM/Rc^2$. Since $T:2\pi = \Delta t : \delta\theta$, $\delta\theta = \pi GM/Rc^2$, one sixth of general relativity result[2].

## 2. Lense-Thirring precession

First deduced by de Sitter [3] in the equatorial plane of the central body, source of the Kerr field, has been then studied by Thirring and Lense [4, p. 365] in the weak field case for all inclinations. Central rotating mass drags the spacetime and its fades far from rotating mass, like a peppercorn deepen in a layer of honey is dragged by the rotation of a nearby ball immersed in the honey, but there is no friction between rotating mass and local spacetime [5]. The Lense-Thirring precession frequency is $\omega_{LT} = GI/2c^2R^3 \cdot [3\mathbf{R}/R^2(\boldsymbol{\omega} \cdot \mathbf{R}) - \boldsymbol{\omega}]$, where the vector $\mathbf{R}$ is the distance from the center of the rotating sphere, of radius r, momentum of inertia I and angular velocity $\boldsymbol{\omega}$. Note that $\omega_{LT} \propto I\omega/R^3$.

Table 1. Lense-Thirring precessions of perihelia in comparison. Data calculated with a momentum of inertia I=2/5MR², valid for homogeneous spheres. For Jupiter's satellite Amalthea, which has an orbit of 12 hours, 2994 arcsecs correspond to 100s of time delay in a century [Thirring & Lense, reprinted in 4].

| Orbiting body | $\omega$ [rad/s] | I [kg·m²/10⁴⁴] | distance [m] | Precession[arcsec/cy] |
|---|---|---|---|---|
| Mercury | $2.7 \cdot 10^{-6}$ | $3.92 \cdot 10^7$ | $5 \cdot 10^{10}$ | -0.0128 |
| Amalthea | $1.7 \cdot 10^{-4}$ | 403.28 | $1.81 \cdot 10^8$ | -2994 |
| Moon | $7.3 \cdot 10^{-5}$ | $9.76 \cdot 10^{-3}$ | $3.84 \cdot 10^8$ | $-1.9 \cdot 10^{-4}$ |
| GP-B | $7.3 \cdot 10^{-5}$ | $9.76 \cdot 10^{-3}$ | $7 \cdot 10^6$ | -3.16 /+6.32 in polar orbit |

### 3. Pugh-Schiff precession

It is the Lense-Thirring precession of an orbiting spinning vector. The Lense-Thirring field acts upon an orbiting gyroscope with a torque $\omega_{LT}$ counter-rotating with respect to the central body $\omega$ in the equatorial plane and co-rotating above the poles. Around the Earth, as Schiff explains: "at the poles there is a tendency for the metric to rotate with the Earth. At the equator we note that the gravitational field, and hence the dragging of the metric, falls off with increasing radial distance. If, then, we imagine the gyroscope oriented so that its axis is perpendicular to the Earth, the side of the gyroscope nearest the earth is dragged with the Earth more than the side away from the Earth, so that the spin precesses in the opposite direction to the rotation of the Earth" [4, p. 436]. Therefore the orbiting spin **s** precesses with a frequency d**s**/dt= $\omega_{LT} \wedge$ **s**. The precessional period for the Earth's spin due to Lense-Thirring is $T_{LT}=2\pi/\omega_{LT}=273 \cdot 10^9$ years.

### 4. Orbits in strong Kerr fields

Since rotating energy gravitates, orbits in Kerr field does no longer lie in a plane according to first Kepler's law. It occurs because of the vectorial composition in the term [3**R**/R²($\omega \cdot$**R**)-$\omega$] of Lense-Thirring angular velocity vector. For polar orbits $\omega_{LT}$ is directed along $\omega$ and orbital angular momentum **L** will precess around $\omega$ leading to a spherical orbit. In strong fields the linear approximation under which previous formulae have been deduced no longer applies and orbits, or equivalently said bound geodetics, form complicate patterns ranging from maximum latitudes (+ and -) to the equatorial plane [Wilkins in 4, p. 469].

### 5. Geodetic precession

Parallel transport (of a constant spin vector) in curved spacetime along a geodetic line (free fall→orbit) generates a precession with respect to a fixed reference. Conversely Thomas precession occurs in flat spacetime (special relativity) with accelerated bodies (non geodetic motion). Thomas precession in General Relativity occurs when additional non gravitational strengths deviate the body from geodetic motion. Rewriting de Sitter precession formula as coupling between angular momenta (spin and orbital), the formula for this precession is $dS/dt = 3/2 \cdot GM_\odot/mc^2R^3 \, (L \wedge S)$ where **S** is the orbital momentum of the Moon, or a constant spinning vector, $L = m \cdot v \wedge R$ is the orbital momentum of the Earth-Moon system of mass m, or that one of the body carrying the constant spinning vector as in the case of GP-B experiment. $M_\odot$ is the solar mass and R the distance from the Sun. This precession is along the direction of the motion and for $L_{Moon}$ is $\Omega_{dS} = 3/2 \cdot GM_\odot/c^2R^2\sqrt{(GM_\odot/R)} \sim 0.0192$ arcsecs per year, already measured within 2% of accuracy by Bertotti et al. [4, p. 497]. The precessional period for Earth's spin is $T_{dS} = 2\pi/\omega_{dS} = 6.7 \cdot 10^7$ years [$5 \cdot 10^7$ in Corinaldesi and Papapetrou in 4, p. 411], while the equinox precession due to lunar and solar coupling with the Earth's quadrupole has T=26000 years. In GP-B experiment 4 gyroscopes have spinning axes oriented in order to measure Lense-Thirring and de Sitter torques [Everitt in 4, p. 439]. It is $\omega_{dS} \propto \sqrt{(M^3/R^5)}$, thence for GP-B satellite around Earth $\Omega_{dS} = 6.6$ arcsecs per year are calculated.

### 6. Classical quadrupole contribution

Around Earth satellite orbits are subjected to the quadrupole precession $\Omega q \propto -(3/2)\omega_0 [R_\oplus/a/(1-e^2)]^2 J_{2\oplus} \cdot \cos(i) \cdot n_\oplus$ with i inclination; $\omega_0$ satellite mean motion and $J_{2\oplus} = -Q^{33}/2MR^3_\oplus$ adimensional parameter for quadrupole momentum; with $Q^{TM}$ quadrupole tensor. Note that Inertia tensor $I = \text{diag}(I_1, I_1, I)$ with $I_1 = I - J_2 \cdot MR^2$. $J_{2\oplus} = -1.083 \cdot 10^{-3}$ and $J_{2\odot} \approx 10^{-7}$. Quadrupole precession vanishes for i=90°, for polar orbit, this is the reason for Van Patten and Everitt's choice of polar orbits in their experiment with twin satellites [in 4, p. 487]. Adler and Silbergleit calculated quadrupole corrections to relativistic torques [4, p. 152]: $\Omega_{dS} = 3GMv/2r^2 \cdot [1 - 9/8 \, J_2 \cdot (R/R_\oplus)^2]$ and $\Omega_{LT} = GI\omega/2R^3 \cdot [1 + 9/8 \cdot J_2 \cdot (R/R_\oplus)^2 \cdot (1 - 88/147 \cdot MR^2/I)]$ where R is the orbital radius and $R_\oplus$ the Earth's one. Those corrections are both of 0.1% of unperturbed values: GP-B expected accuracy will be sufficient to detect this correction only for the de Sitter precession.

**References**

1. Einstein, A., Preuss. Akad. Wiss. Berlin, **47**, 831 (1915).
2. Goldstein, H., *Classical Mechanics*, Addison-Wesley, MA-USA (1965).
3. de Sitter, W., Mon. Not. R. Astron. Soc. **76**, 699 (1916).
4. R. Ruffini and C. Sigismondi editors, *Nonlinear Gravitodynamics, the Lense-Thirring effect,* World Scientific pub., Singapore (2003).
5. S. K. Range, *Curriculum Connections to Gravity Probe-B,* http://einstein.stanford.edu (2004).
6. M. H. Soffel, *Relativity in Astrometry,* Springer-Verlag, Berlin (1989).

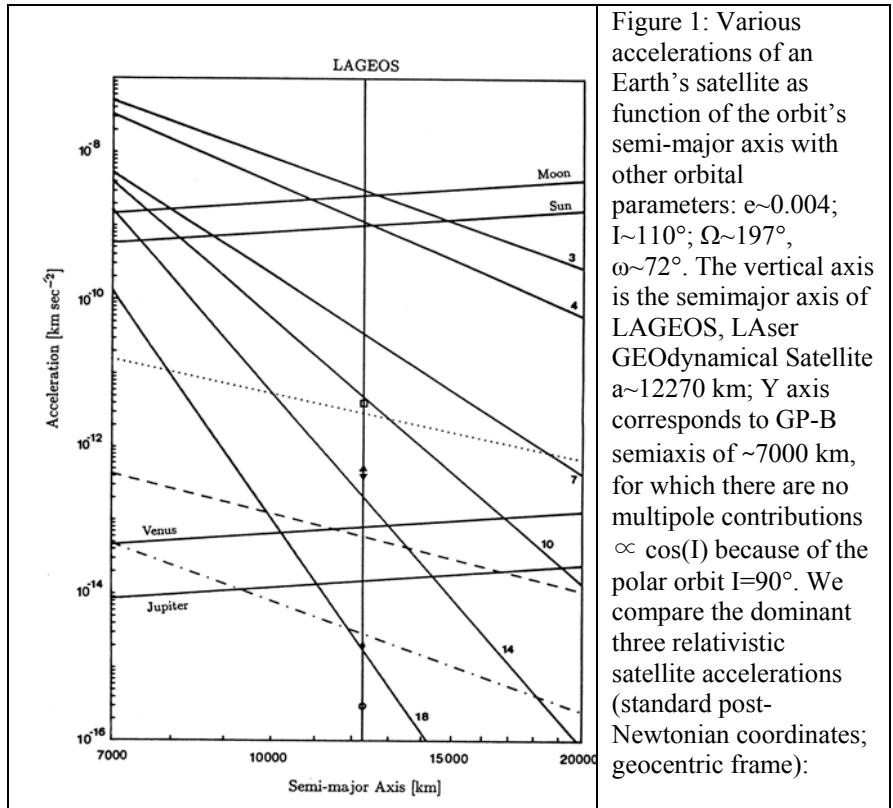

Figure 1: Various accelerations of an Earth's satellite as function of the orbit's semi-major axis with other orbital parameters: e~0.004; I~110°; Ω~197°, ω~72°. The vertical axis is the semimajor axis of LAGEOS, LAser GEOdynamical Satellite a~12270 km; Y axis corresponds to GP-B semiaxis of ~7000 km, for which there are no multipole contributions ∝ cos(I) because of the polar orbit I=90°. We compare the dominant three relativistic satellite accelerations (standard post-Newtonian coordinates; geocentric frame):

i) dotted curve: the contribution from the post-Newtonian spherical field of the Earth. The post- Newtonian spherical field of the Earth leads to the well known Einstein perihelion precession, that was first detected in the orbit of Mercury, and which in LAGEOS case is $\Delta\Omega$~3 arcsec/year. At LAGEOS distances Einstein precession is of the same amount of the geodetic precession; ii) dashed curve: the Lense-Thirring acceleration which induces an additional perihelion precession and a secular drift of the nodes of the same order of magnitude; iii) dotted-dash curve: the relativistic acceleration due to the oblateness of the Earth with a set of other accelerations as a function of the semi-major axis of the satellite's orbit (all other parameters as those of LAGEOS). Here the contributions from the various zonal multipole moments are indicated by the corresponding index, they become steeper at higher numbers. The tidal accelerations caused by the Moon, Sun, Venus and Jupiter are indicated by the names of the celestial bodies. For the LAGEOS orbit we also included estimates of the direct solar radiation pressure (□), the Earth's albedo (△), infrared pressure (▽) and the charged particle (◇) and neutral particle (○) drag. Note that at this level of accuracy the orbit analysis and prediction face the problem of modeling accelerations as small as ~$10^{-15}$ km·s$^{-2}$. Figure reprinted with kind permission of the publisher from M. H. Soffel 1989 [6].

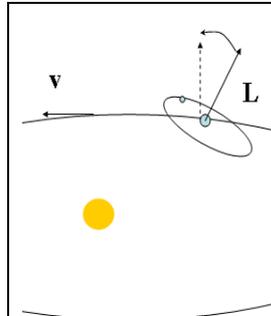

Figure 2: Geodetic precession. The case of Earth-Moon system. After one revolution the orbital momentum **L** changes because parallel transport in curved spacetime.

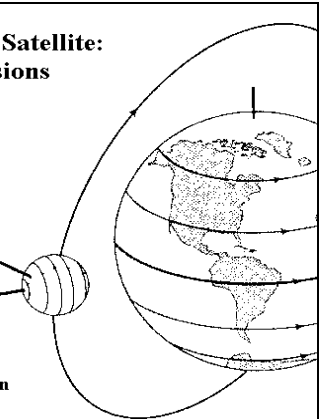

Figure 3: Relativistic precessions acting on each of GP-B four gyroscopes. Their axes are oriented towards a radio star IM Peg (HR8703) whose proper motion is very well known with VLBI because this star is near a distant quasar.

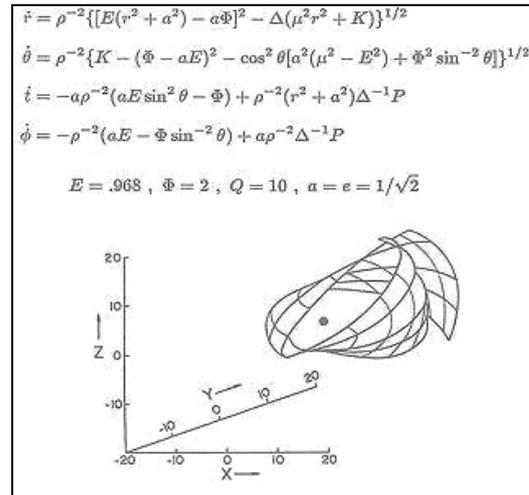

Figure 4: Orbits in strong Kerr-Neumann fields. Motion of uncharged cloud of particles corotating about an extreme spinning black hole. The orbits are stable; equations and constants of motions are given in the figure. The vertical lines indicate isochronous points as seen from infinity. From Johnston and Ruffini (ref. 4 p. 478). It become the ICRA logo.

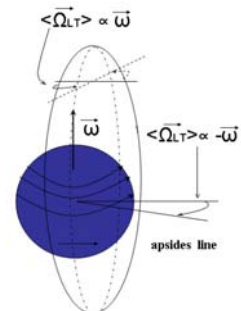

Figure 5: Lense-Thirring torque: the spacetime above the poles rotates with them, while the rotation of the Earth drags more strongly the spacetime where is the side of a spinning body closer to the equator than the other making it counter rotate with respect to the Earth. The effect of dragging on the orbits' apsides is shown.